\documentclass[%
 reprint,
 amsmath,amssymb,
 aps,
]{revtex4-2}

\usepackage{graphicx}
\usepackage{dcolumn}
\usepackage{bm}
\usepackage{xcolor}

\begin{document}

\preprint{APS/123-QED}

\title{Superradiant droplet emission from parametrically excited cavities}

\author{Valeri Frumkin}
\thanks {These authors contributed equally to this work}
\affiliation{Department of Mathematics, Massachusetts Institute of Technology.}

\author{Konstantinos Papatryfonos}
\thanks {These authors contributed equally to this work}
\affiliation{Department of Mathematics, Massachusetts Institute of Technology} 
\affiliation{Gulliver UMR CNRS 7083, ESPCI Paris, Université PSL, 75005 Paris, France.}

\author{John W. M. Bush}
 \email{bush@math.mit.edu}
\affiliation{Department of Mathematics, Massachusetts Institute of Technology.}

\begin{abstract}
Superradiance occurs when a collection of atoms exhibits cooperative, spontaneous emission of photons at a rate that exceeds that of its component parts.
Here, we reveal a similar phenomenon in a hydrodynamic system consisting of a pair of vibrationally-excited cavities, coupled through their common wavefield, that spontaneously emit droplets via interfacial fracture. We show that the droplet emission rate of two coupled cavities is higher than the emission rate of two isolated cavities. We further show that the amplified emission rate varies sinusoidally with distance between the cavities, thus demonstrating a hydrodynamic phenomenon that captures the essential features of superradiance in optical systems. 
\end{abstract}

\maketitle

When a group of $N$ quantum emitters ({\it e.g.} excited atoms) interact coherently with a common electromagnetic field, they may collectively emit photons at a rate that is amplified relative to that corresponding to the sum of the individual emitters. In quantum optics, this phenomenon is known as superradiance \citep{dicke_coherence_1954}, an effect of both fundamental and practical interest, with applications in various fields, including quantum information technologies \citep{kalachev_quantum_2007, black_-demand_2005,scully_single_2015}, cryptography \citep{ekert_quantum_1991}, and narrow linewidth lasers \citep{meiser_prospects_2009,bohnet_steady-state_2012,svidzinsky_quantum_2013}. When the separation between the atoms is much smaller than the emission wavelength, superradiance can be understood by picturing each atom as a tiny antenna emitting electromagnetic waves \citep{scully_super_2009}, the result being photon emission at a rate proportional to $N^2$ \citep{zakowicz_collective_1974,tralle_induced_2014}. A more puzzling type of superradiance occurs when the separation between the atoms is comparable to the emission wavelength.  
According to the quantum mechanical description, this type of superradiance occurs when at each absorption event, a single photon is stored in a cloud of N atoms of the same kind. The atoms interact with each other through the electromagnetic field, creating collective non-separable states that can radiate the photon faster or slower than if the photon were stored in a single atom, corresponding to super– or subradiant emission states, respectively \citep{scully_super_2009, gross_superradiance_1982, solano_super-radiance_2017}. 
An important feature of this second type of superradiance is the sinusoidal modulation of the spontaneous emission rate with distance between emitters \citep{power_effect_1967}, which was first demonstrated using a pair of trapped ions whose separation distance, $d$, was varied gradually \citep{devoe_observation_1996}. Experiments revealed sinusoidal oscillations of the spontaneous emission rate $\Gamma(d)$ of the two-ion crystal, in accord with detailed quantum mechanical theoretical analysis \citep{makarov_spontaneous_2003}. This type of superradiance was first considered to be a purely quantum phenomenon, but has since been rationalized in terms of classical electromagnetic theory \cite{tanji-suzuki_interaction_2011}.

\begin{figure} [t]
\noindent \begin{centering}
\hspace*{-0.5cm}
\includegraphics[width=21pc]{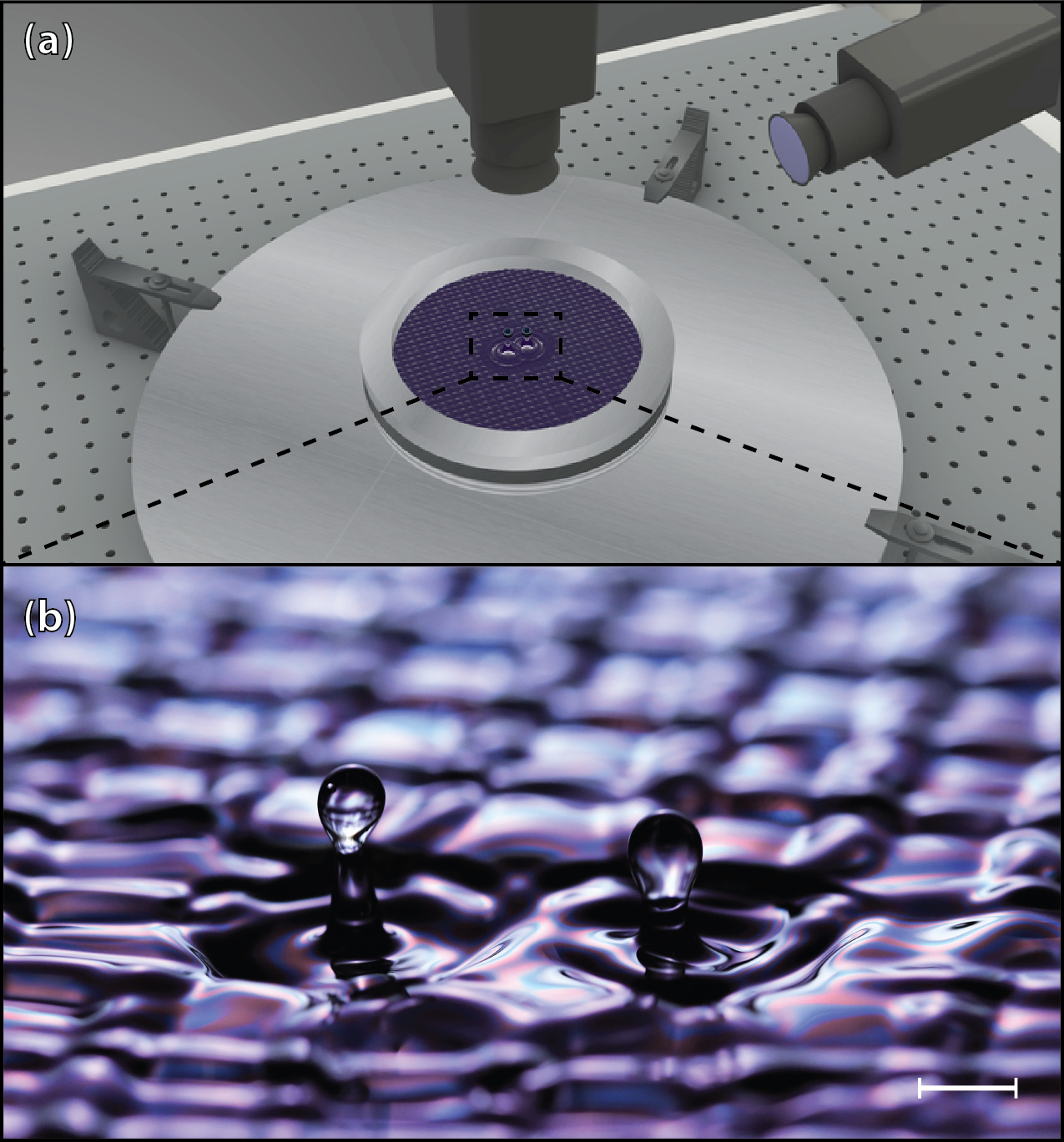}
\par\end{centering}
    \caption{The experimental setup (see also Supplementary Fig. 1). (a) A schematic illustration of a circular bath with two cavities spanned by a thin layer of fluorinated oil. The bath is vertically oscillated by an electromagnetic shaker, resulting in the emission of droplets from the two cavities. (b) A rare generation event in which two droplets are about to be created simultaneously. Scale bar, 3 mm.}
    \label{fig:1}
\end{figure}

Fluid mechanics has produced laboratory-scale physical analogs for phenomena as disparate as the wave nature of light \citep{young_i_1804}, black holes \citep{unruh_experimental_1981}, the Casimir effect \citep{denardo_water_2009} and the Aharonov-Bohm effect \citep{berry_wavefront_1980}. The relatively recent discovery of a pilot-wave hydrodynamic system \citep{couder_single-particle_2006} 
has led to a new class of hydrodynamic quantum analogs~\citep{bush_hydrodynamic_2020} that includes analogs of quantized orbital states \citep{fort_path-memory_2010, perrard_self-organization_2014}, quantum corrals \citep{harris_wavelike_2013, saenz_statistical_2018}, Friedel oscillations \citep{saenz_hydrodynamic_2020} and spin lattices \citep{saenz_emergent_2021}.
We here report a hydrodynamic phenomenon that is in many ways analogous to that of superradiance. We consider a system of vibrationally excited hydrodynamic cavities that spontaneously emit droplets via interfacial fracture. The cavities are deep circular wells spanned by a thin layer of oil that allows for their coupling through a common wavefield (see Fig. 1). We demonstrate that the wavefield in each cavity is influenced by the presence of its neighbor. Specifically, the neighboring cavity may amplify the local oscillation amplitude, resulting in an increased chance of interfacial fracture and thus an amplified droplet emission rate.

Figure 1 shows a schematic representation of our experimental set-up. A bath of fluorinated oil has two $6$-mm-deep circular wells that serve as hydrodynamic cavities. The cavities, each with diameter $7$ mm, are separated by a center-to-center distance $d$ that is varied between experiments, from $8 $ mm to $12$ mm, in $0.5$ mm increments. In the shallow layer spanning the wells, the depth is $0.75\pm 0.05$ mm. The system is subjected to vertical vibration by an electromagnetic shaker with forcing $F(t)=\gamma\cos(2\pi ft)$, where $\gamma=1.75 g$ and $f=39\,\, Hz$ are the peak driving acceleration and frequency, respectively. A more detailed description of the experimental setup is provided in the supplementary information (SI).

A liquid bath of uniform depth subject to vertical vibration at a fixed frequency, undergoes two critical transitions as the driving amplitude is increased progressively. The first transition occurs as the vibrational acceleration, $\gamma=4\pi^2f^2A$, where $A$ is the vibration amplitude, is increased beyond the Faraday threshold, $\gamma_F$, at which point the initially flat free surface destabilizes into a pattern of standing Faraday waves \citep{faraday_peculiar_1837}. As the driving amplitude is increased further, the stabilizing influence of surface tension is exceeded by the destabilizing inertial forces associated with the bath vibration, and the interfacial fracture threshold, $\gamma_B$, is crossed. Above this threshold, the Faraday waves break spontaneously, and millimetric droplets are emitted from the free surface in an irregular fashion \citep{goodridge_viscous_1997,goodridge_breaking_1999}. Importantly, for shallow layers, both $\gamma_F$  and $\gamma_B$  depend strongly on the local depth of the liquid. We thus define $\gamma_F^c$ and $\gamma_F^s$ to be the Faraday threshold above the cavities and the shallow region, respectively, and likewise for $\gamma_B$. 

\begin{figure} [t!]
\noindent \begin{centering}
\hspace*{-0.5cm}
\includegraphics[width=21pc]{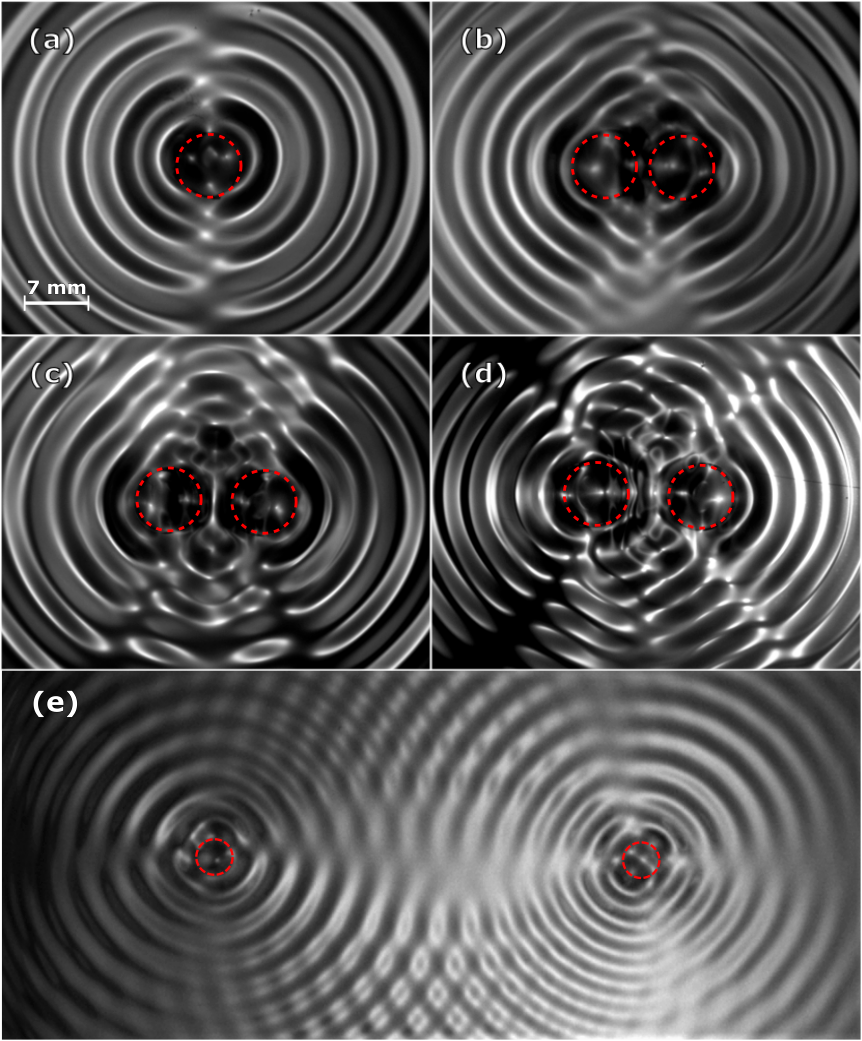}
\par\end{centering}
    \caption{Images of the instantaneous wavefield generated by the cavities (red circles). (a)  A single cavity. Two cavities with center-to-center distances of (b) $d=8$ mm. (c) $d=10.5$ mm. (d)  $d=12$ mm. (e) $d=87$ mm.}
    \label{fig:2}
\end{figure}

With the increase of the driving acceleration, $\gamma$, our variable-depth system undergoes the following evolution. First, as the acceleration crosses $\gamma_F^c$, Faraday waves appear above the cavities and propagate some distance into the surrounding shallow region. When $\gamma>\gamma_F^s$, Faraday waves emerge over the entirety of the bath surface, but are most vigorous above the cavities. Figure 2 illustrates the instantaneous wavefield near the cavities when $\gamma_F^s  < \gamma <  \gamma_B^c$. Figure 2a shows the wavefield of a single cavity, whereas figures 2b-d show the two-cavity wavefield for three different values of the center-to-center separation distance, $d=8,\,10.5,\,12\,$ mm, respectively. Figure 2e depicts the resulting wavefield for two distant cavities, with $d=87$ mm. Notably, even at such large separation distances, the perturbation wavefield can reach the other cavity, allowing for long-range interactions. The perturbation wavefield, recorded near the frequency of the most unstable Faraday mode, $f/2$, is shown in Supplementary Movie 1.

When the acceleration is increased beyond the interfacial breaking threshold of the cavities,  $\gamma_B^c<\gamma<\gamma_B^s$, droplet emission sets in. Supplementary Movie 2 shows the spontaneous droplet emission from a pair of hydrodynamic cavities. The emission events occur unpredictably, as indicated by Fourier analysis shown in the SI, but arise exclusively within the cavities. We define a spontaneous emission rate, $\Gamma$, for the combined two-cavity system, as the average number of emission events per second, and the anomalous emission rate, $\Gamma_N(d)=\left(\Gamma(d)-2\Gamma_0\right)/2\Gamma_0$, where $\Gamma_0=1.47\,s^{-1}$, is the measured emission rate of a single cavity in isolation. 

\begin{figure}[t!]
\noindent \begin{centering}
\hspace*{-0.5cm}
\includegraphics[width=21pc]{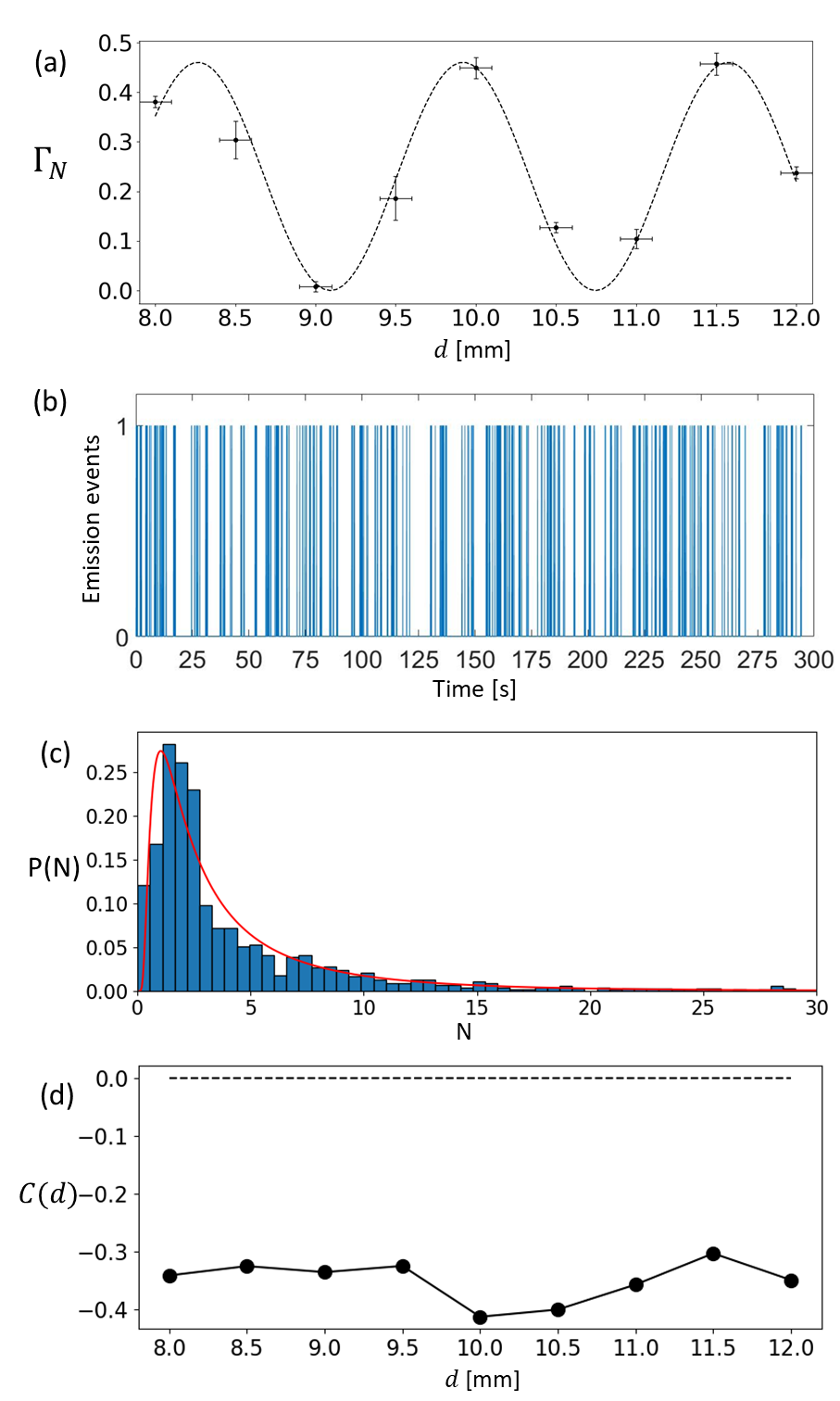}
\par\end{centering}
    \caption{Experimental measurements of the droplet emission rate: (a) Dependence of the anomalous emission rate, $\Gamma_N(d) = \left(\Gamma(d)-2\Gamma_0\right)/2\Gamma_0$ (black dots) on center-to-center intercavity distance $d$. Each data point represents an average over a time interval of $300$ seconds, corresponding to roughly $950-1300$ droplet emission events.
    The dashed curve represents  $A \cos^2(2kd)$, with $A=1.36$ and $k=0.85$ mm$^{-1}$ being the experimentally measured Faraday wave number in the vicinity of the cavities.
    (b) Time series of the emission events from a single cavity over a 300-second interval indicates the unpredictability of a single emission event. (c) Histogram of the inter-emission time intervals and its comparison to the first passage time model \cite{folks_inverse_1978}. The red curve represents an inverse Gaussian distribution as given by Eq. $(2)$.
    (d) Measured correlation, $C$, of the drop emission event in the two cavities as a function of their separation distance $d$. The upper dashed lined represents two uncorrelated cavities.}
    \label{fig:3}
\end{figure}

In Figure 3(a), we present our experimental measurements of the dependence of the anomalous emission rate $\Gamma_N(d)$ on the separation distance $d$.  An amplification of up to $46 \%$ relative to $2\Gamma_0$ is evident.
One expects the emission rate to be a function of the kinetic energy of the fluid inside the
cavity. For an isolated cavity, the
average kinetic energy per oscillation period is constant, corresponding to a steady emission rate, $\Gamma_0$. In the presence of a second, coupled cavity, the wave-field inside each cavity is effected by its neighbor, and the average kinetic energy per oscillation period will depend on the distance between the two. As in the trapped ion pair experiment \citep{devoe_observation_1996}, the amplified emission rate of our two-cavity system oscillates sinusoidally as a function of the distance between the cavities. The observed oscillatory behaviour shows that the probability of the emission events is affected by the interference between the waves generated by the individual cavities. 
The dashed curve in Fig. 3(a) represents a simple fit for the anomalous emission rate, $\Gamma_N(x)=A\cos^2(2kd)$ for $A=1.36$, and $k=2\pi/\lambda$, with $\lambda=6.60\pm 0.05$ mm being the experimentally measured wavelength of the Faraday waves in the vicinity of the cavities. 
Owing to experimental limitations, we did not systematically characterize the decay of the anomalous emission rate with increasing separation distance, $d$. Nevertheless, we note that the influence of the neighbouring cavity’s wavefield will decay with $d$ due to viscous damping, as will the anomalous emission rate.

Figure 3(b) depicts the time dependence of the emission events from a single cavity, showing the unpredictability of a single emission event, as is confirmed by the FFT analysis presented in the SI. While the highly non-linear and chaotic nature of the emission events makes direct modeling of the emission phenomenon a daunting task, we proceed by demonstrating that the problem lends itself to a stochastic approach. 
Let $X_t$ be a stochastic process representing the maximal wavefield amplitude inside the cavities, with $X_0=0$ representing the initially flat state. Between each two consecutive emission events, we assume that the maximal wavefield amplitude stochastically oscillates around some mean value $\mu(t)$ that grows in time as a result of the resonant interaction between the external forcing and the cavities. Eventually, the maximal amplitude crosses a threshold $\alpha$, resulting in an emission of a droplet, after which $X_t$ relaxes back to $X_0=0$ and the process starts over.
Thus, one may write:
\begin{equation}
    X_t=\nu t+\sigma W_t,\,\,\,\,\,\,\nu,\sigma>0
\end{equation}
where $\nu t$ is a stochastic drift representing the increase in $\mu(t)$ between consecutive emission events, and
$W_t$ is a Wiener process with an amplitude $\sigma$, representing the stochastic oscillations of $X_t$ about $\mu(t)$. The emission process can thus be modeled as a first passage time problem, where we seek to find the first time that $X_t$ reaches the critical value $\alpha$, at which point an emission event occurs. For a Wiener process with a stochastic drift, the probability density function for the first passage time is given by the inverse Gaussian distribution \citep{folks_inverse_1978}:
\begin{equation}
    P(T_{\alpha})=\sqrt{\frac{\lambda}{2\pi T^3}}\exp\left(-\frac{\lambda(T-\mu)^2}{2\mu^2 T}\right)
\end{equation}
where $T_{\alpha}$ is the first time $X_t$ crosses the threshold value $\alpha$, $\lambda=\frac{\alpha}{\nu}$ is the shape parameter, and $\mu=\frac{\alpha}{\nu}$ is the mean value of $T_{\alpha}$.
Fig. 2c presents a histogram of the experimentallt measured inter-emission time intervals for the case $d=12$ mm, showing good agreement with Eq. $(2)$ when $\lambda=3.3$ is used for the shape parameter.

The mechanism responsible for the superradiant emission of droplets is the wave coupling between the two cavities. We quantify this coupling by measuring the correlations between the emission events in the two-cavity system, as detailed in the SI. We see that the two cavities are strongly anti-correlated, with the correlation values varying from $C=-0.30$ (for $d=11.5$ mm) to $C=-0.41$ (for $d=10$ mm). Figure 4 shows the measured correlation, $C$, of the drop emission event in the two cavities as a function of their separation distance $d$.
These anticorrelations, together with the amplification of the combined emission rate, suggest that the two cavity system cannot be factored into distinct states, as the probabilities of emission events in the two cavities are coupled. Acting on one of the cavities of this coupled system, by, for example, changing its position or depth, would affect the emission rate of its neighbouring cavity. The possibility thus arises of altering the system's global emission rate by a local operation on one of its individual components, thereby creating a new platform for probabilistic computational operations in fluid mechanics.

It is also worth considering the relation between the system introduced here and pilot-wave hydrodynamics \citep{bush_hydrodynamic_2020}.  
In the latter, the notion of an analog photon is more nebulous: when the system jumps between quantized states (e.g. the walking droplet transitions from one orbit to another), energy is dumped into or extracted from the bath.
In the system considered here, droplets are generated by breaking waves, their appearance representing a discrete transition event, an analog of photon emission from an excited state. We note that in our current experiments, we used fluorinated oil in order to facilitate the rapid reabsorption of the emitted droplets into the bath. However, this reabsorbtion can be minimized by using a relatively low density silicon oil, in which case the generated particles may persist on the surface, bounce and self-propel, thereby providing a possible link between pilot-wave hydrodynamics and the new class of analog systems established here.

We have introduced a new hydrodynamic system that shares several key features with the phenomenon of superradiance, as it appears in quantum optics. In addition to the amplification of the emission rate typically associated with superradiance, our system exhibits sinusoidal dependence of the amplified emission rate on separation distance (see Fig. 3b), which in our system arises from classical wave interference.

In addition to the aforementioned similarities, we identify several notable differences between optical superradiance and our hydrodynamic analogy thereof. First, in our experiments we did not characterize the structure of the energy levels as would potentially be prescribed by the size and kinetic energy of the emitted droplets, or the transition rates associated with these levels, both of which are well characterized in the quantum mechanical system. Second, in our experiments we did not observe subradiant droplet emission. We believe that the later is due to the chosen cavity geometry precluding the possibility of robust destructive interference. Specifically, in order to support a single oscillatory mode in each cavity, the liquid bath would need to be strongly driven at a frequency of approximately $15$ Hz, which was unreachable with our current setup. Driving the system at $39$ Hz excited higher harmonics inside the cavities, yielding a complex 2-D wavefield which could not be canceled by the ordered wavefield in the shallow inter-cavity region.  
We note that in the quantum mechanical case, while superradiance is readily observed in a wide variety of optical systems, sub-radiance is relatively rare \citep{guerin_subradiance_2016}.

Another interesting comparison can be made between the generation of droplets in the hydrodynamic system and emission of photons in the quantum mechanical one. Both processes represent dissipation mechanisms, the rates of which depend nonlinearly on the amplitude of the relevant field, giving rise to the enhanced emission rate. In the hydrodynamic case, the probability of random discrete events, specifically drop ejection, is prescribed by a continuous wavefield resulting from two interfering sources. This statistical behavior is reminiscent of the way probabilities of outcomes are obtained, in accordance with Born's rule in the standard quantum theory. It would thus be of interest to explore this similarity further.

Finally, our study suggests that droplet creation through interfacial fracture may provide a valuable new platform for exploring hydrodynamic analogs of collective particle emission phenomena, and further expand the field of hydrodynamic quantum analogs.

\bibliographystyle{ieeetr}
\bibliography{Superradiance.bib}

\begin{thebibliography}{10}

\bibitem{dicke_coherence_1954}
R.~H. Dicke, ``Coherence in {Spontaneous} {Radiation} {Processes},'' {\em
  Physical Review}, vol.~93, pp.~99--110, Jan. 1954.

\bibitem{kalachev_quantum_2007}
A.~Kalachev, ``Quantum storage on subradiant states in an extended atomic
  ensemble,'' {\em Physical Review A}, vol.~76, p.~043812, Oct. 2007.
\newblock Publisher: American Physical Society.

\bibitem{black_-demand_2005}
A.~T. Black, J.~K. Thompson, and V.~Vuletić, ``On-{Demand} {Superradiant}
  {Conversion} of {Atomic} {Spin} {Gratings} into {Single} {Photons} with
  {High} {Efficiency},'' {\em Physical Review Letters}, vol.~95, p.~133601,
  Sept. 2005.

\bibitem{scully_single_2015}
M.~O. Scully, ``Single {Photon} {Subradiance}: {Quantum} {Control} of
  {Spontaneous} {Emission} and {Ultrafast} {Readout},'' {\em Physical Review
  Letters}, vol.~115, p.~243602, Dec. 2015.

\bibitem{ekert_quantum_1991}
A.~K. Ekert, ``Quantum cryptography based on {Bell}'s theorem,'' {\em Physical
  Review Letters}, vol.~67, pp.~661--663, Aug. 1991.

\bibitem{meiser_prospects_2009}
D.~Meiser, J.~Ye, D.~R. Carlson, and M.~J. Holland, ``Prospects for a
  {Millihertz}-{Linewidth} {Laser},'' {\em Physical Review Letters}, vol.~102,
  p.~163601, Apr. 2009.

\bibitem{bohnet_steady-state_2012}
J.~G. Bohnet, Z.~Chen, J.~M. Weiner, D.~Meiser, M.~J. Holland, and J.~K.
  Thompson, ``A steady-state superradiant laser with less than one intracavity
  photon,'' {\em Nature}, vol.~484, pp.~78--81, Apr. 2012.

\bibitem{svidzinsky_quantum_2013}
A.~A. Svidzinsky, L.~Yuan, and M.~O. Scully, ``Quantum {Amplification} by
  {Superradiant} {Emission} of {Radiation},'' {\em Physical Review X}, vol.~3,
  p.~041001, Oct. 2013.

\bibitem{scully_super_2009}
M.~O. Scully and A.~A. Svidzinsky, ``The {Super} of {Superradiance},'' {\em
  Science}, vol.~325, pp.~1510--1511, Sept. 2009.

\bibitem{zakowicz_collective_1974}
W.~Zakowicz and K.~Rzazewski, ``Collective radiation by harmonic oscillators,''
  {\em Journal of Physics A: Mathematical, Nuclear and General}, vol.~7,
  pp.~869--880, May 1974.

\bibitem{tralle_induced_2014}
I.~Tralle and P.~Zieba, ``Induced {N2}-cooperative phenomenon in an ensemble of
  the nonlinear coupled oscillators,'' {\em Physics Letters A}, vol.~378,
  pp.~1364--1368, Apr. 2014.

\bibitem{gross_superradiance_1982}
M.~Gross and S.~Haroche, ``Superradiance: {An} essay on the theory of
  collective spontaneous emission,'' {\em Physics Reports}, vol.~93,
  pp.~301--396, Dec. 1982.

\bibitem{solano_super-radiance_2017}
P.~Solano, P.~Barberis-Blostein, F.~K. Fatemi, L.~A. Orozco, and S.~L. Rolston,
  ``Super-radiance reveals infinite-range dipole interactions through a
  nanofiber,'' {\em Nature Communications}, vol.~8, p.~1857, Nov. 2017.

\bibitem{power_effect_1967}
E.~A. Power, ``Effect on the {Lifetime} of an {Atom} {Undergoing} a {Dipole}
  {Transition} {Due} to the {Presence} of a {Resonating} {Atom},'' {\em The
  Journal of Chemical Physics}, vol.~46, pp.~4297--4298, June 1967.
\newblock Publisher: American Institute of Physics.

\bibitem{devoe_observation_1996}
R.~G. DeVoe and R.~G. Brewer, ``Observation of {Superradiant} and {Subradiant}
  {Spontaneous} {Emission} of {Two} {Trapped} {Ions},'' {\em Physical Review
  Letters}, vol.~76, pp.~2049--2052, Mar. 1996.

\bibitem{makarov_spontaneous_2003}
A.~A. Makarov and V.~S. Letokhov, ``Spontaneous decay in a system of two
  spatially separated atoms ({One}-dimensional case),'' {\em Journal of
  Experimental and Theoretical Physics}, vol.~97, pp.~688--701, Oct. 2003.

\bibitem{tanji-suzuki_interaction_2011}
H.~Tanji-Suzuki, I.~D. Leroux, M.~H. Schleier-Smith, M.~Cetina, A.~T. Grier,
  J.~Simon, and V.~Vuletic, ``Interaction between {Atomic} {Ensembles} and
  {Optical} {Resonators}: {Classical} {Description},'' Nov. 2011.
\newblock Number: arXiv:1104.3594 arXiv:1104.3594 [quant-ph].

\bibitem{young_i_1804}
T.~Young, ``I. {The} {Bakerian} {Lecture}. {Experiments} and calculations
  relative to physical optics,'' {\em Philosophical Transactions of the Royal
  Society of London}, vol.~94, pp.~1--16, Jan. 1804.

\bibitem{unruh_experimental_1981}
W.~G. Unruh, ``Experimental {Black}-{Hole} {Evaporation}?,'' {\em Physical
  Review Letters}, vol.~46, pp.~1351--1353, May 1981.

\bibitem{denardo_water_2009}
B.~C. Denardo, J.~J. Puda, and A.~Larraza, ``A water wave analog of the
  {Casimir} effect,'' {\em American Journal of Physics}, vol.~77,
  pp.~1095--1101, Dec. 2009.

\bibitem{berry_wavefront_1980}
M.~V. Berry, R.~G. Chambers, M.~D. Large, C.~Upstill, and J.~C. Walmsley,
  ``Wavefront dislocations in the {Aharonov}-{Bohm} effect and its water wave
  analogue,'' {\em European Journal of Physics}, vol.~1, pp.~154--162, July
  1980.

\bibitem{couder_single-particle_2006}
Y.~Couder and E.~Fort, ``Single-{Particle} {Diffraction} and {Interference} at
  a {Macroscopic} {Scale},'' {\em Physical Review Letters}, vol.~97, p.~154101,
  Oct. 2006.

\bibitem{bush_hydrodynamic_2020}
J.~W.~M. Bush and A.~U. Oza, ``Hydrodynamic quantum analogs,'' {\em Reports on
  Progress in Physics}, vol.~84, p.~017001, Dec. 2020.

\bibitem{fort_path-memory_2010}
E.~Fort, A.~Eddi, A.~Boudaoud, J.~Moukhtar, and Y.~Couder, ``Path-memory
  induced quantization of classical orbits,'' {\em Proceedings of the National
  Academy of Sciences of the United States of America}, vol.~107,
  pp.~17515--17520, Oct. 2010.

\bibitem{perrard_self-organization_2014}
S.~Perrard, M.~Labousse, M.~Miskin, E.~Fort, and Y.~Couder, ``Self-organization
  into quantized eigenstates of a classical wave-driven particle,'' {\em Nature
  Communications}, vol.~5, p.~3219, Jan. 2014.

\bibitem{harris_wavelike_2013}
D.~M. Harris, J.~Moukhtar, E.~Fort, Y.~Couder, and J.~W.~M. Bush, ``Wavelike
  statistics from pilot-wave dynamics in a circular corral,'' {\em Physical
  Review E}, vol.~88, p.~011001, July 2013.

\bibitem{saenz_statistical_2018}
P.~J. Sáenz, T.~Cristea-Platon, and J.~W.~M. Bush, ``Statistical projection
  effects in a hydrodynamic pilot-wave system,'' {\em Nature Physics}, vol.~14,
  pp.~315--319, Mar. 2018.

\bibitem{saenz_hydrodynamic_2020}
P.~J. Sáenz, T.~Cristea-Platon, and J.~W.~M. Bush, ``A hydrodynamic analog of
  {Friedel} oscillations,'' {\em Science Advances}, vol.~6, p.~eaay9234, May
  2020.

\bibitem{saenz_emergent_2021}
P.~J. Sáenz, G.~Pucci, S.~E. Turton, A.~Goujon, R.~R. Rosales, J.~Dunkel, and
  J.~W.~M. Bush, ``Emergent order in hydrodynamic spin lattices,'' {\em
  Nature}, vol.~596, pp.~58--62, Aug. 2021.

\bibitem{faraday_peculiar_1837}
M.~Faraday, ``On a peculiar class of acoustical figures; and on certain forms
  assumed by groups of particles upon vibrating elastic surfaces,'' {\em
  Abstracts of the Papers Printed in the Philosophical Transactions of the
  Royal Society of London}, vol.~3, pp.~49--51, Jan. 1837.

\bibitem{goodridge_viscous_1997}
C.~L. Goodridge, W.~T. Shi, H.~G.~E. Hentschel, and D.~P. Lathrop, ``Viscous
  effects in droplet-ejecting capillary waves,'' {\em Physical Review E},
  vol.~56, pp.~472--475, July 1997.

\bibitem{goodridge_breaking_1999}
C.~L. Goodridge, H.~G.~E. Hentschel, and D.~P. Lathrop, ``Breaking {Faraday}
  {Waves}: {Critical} {Slowing} of {Droplet} {Ejection} {Rates},'' {\em
  Physical Review Letters}, vol.~82, pp.~3062--3065, Apr. 1999.

\bibitem{folks_inverse_1978}
J.~L. Folks and R.~S. Chhikara, ``The {Inverse} {Gaussian} {Distribution} and
  its {Statistical} {Application}—{A} {Review},'' {\em Journal of the Royal
  Statistical Society: Series B (Methodological)}, vol.~40, no.~3,
  pp.~263--275, 1978.
\newblock \_eprint:
  https://onlinelibrary.wiley.com/doi/pdf/10.1111/j.2517-6161.1978.tb01039.x.

\bibitem{guerin_subradiance_2016}
W.~Guerin, M.~O. Araújo, and R.~Kaiser, ``Subradiance in a {Large} {Cloud} of
  {Cold} {Atoms},'' {\em Physical Review Letters}, vol.~116, p.~083601, Feb.
  2016.

\end{thebibliography}

\vspace{0.5cm}
\subsection*{Acknowledgments} We thank Masha Bluvshtein for her help with analyzing the data, and Christos Orestis Apostolidis for providing the graphical schematics used in this paper. We also thank Matthieu Labousse and André Nachbin for valuable discussions, as well as Shimon Rubin and Tal Kachman for their constructive comments on the manuscript.
\subsubsection*{Funding}
The authors gratefully acknowledge the financial support of the United States Department of State (V.F.),
the European Union’s Horizon 2020 research and innovation programme under the Marie Sklodowska-Curie project EnHydro, grant agreement No 841417 (K.P.),
and the National Science Foundation grant CMMI-1727565 (J.B.).

\subsubsection*{Competing interests}
The authors declare no competing interests.

\subsubsection*{Data and materials availability}
All data are available in the main text or the supplementary materials.

\end{document}